\documentclass[usenatbib]{mn2e}
\usepackage{amsmath}
\usepackage{amssymb}
\usepackage{graphicx}
\usepackage{hyperref}

\newcommand{\be}{\begin{equation}}
\newcommand{\ee}{\end{equation}}
\newcommand{\beq}{\begin{eqnarray}}
\newcommand{\eeq}{\end{eqnarray}}
\newcommand{\n}{\mathrm{n}}
\newcommand{\cc}{\mathrm{p}}
\newcommand{\LL}{\mathrm{L}}
\newcommand{\RR}{1+\mathcal{R}^2}
\title[Pinned vortex hopping in a neutron star crust]{Pinned vortex hopping in a neutron star crust}
\author[B.Haskell \& A.Melatos]{B.~Haskell \& A.~Melatos\\
School of Physics, The University of Melbourne, Parkville, VIC 3010, Australia}

\begin{document}

\pagerange{\pageref{firstpage}--\pageref{lastpage}} \pubyear{2014}

\maketitle

\label{firstpage}

\begin{abstract}

The motion of superfluid vortices in a neutron star crust is at the heart of most theories of pulsar glitches. Pinning of vortices to ions can decouple the superfluid from the crust and create a reservoir of angular momentum. Sudden large scale unpinning can lead to an observable glitch. In this paper we investigate the scattering of a free vortex off a pinning potential and calculate its mean free path, in order to assess whether unpinned vortices can skip multiple pinning sites and come close enough to their neighbours to trigger avalanches, or whether they simply hop from one pinning site to another giving rise to a more gradual creep. We find that there is a significant range of parameter space in which avalanches can be triggered, thus supporting the hypothesis that they may lie at the origin of pulsar glitches. For realistic values of the pinning force and superfluid drag parameters we find that avalanches are more likely in the higher density regions of the crust where pinning is stronger. Physical differences in stellar parameters, such as mass and temperature, may lead to a switch between creep-like motion and avalanches, explaining the different characteristics of glitching pulsars.

\end{abstract}


\section{Introduction}

At the high densities that characterise Neutron Star (NS) interiors, neutrons are expected to form Cooper pairs and become superfluid \citep{migdal}. This has profound consequences for the dynamics of the system, as the neutrons couple weakly to the `normal' part of the star.  As the normal component of a magnetised NS spins down electromagnetically part of the superfluid may decouple from the bulk of the star and not spin down, building up a lag between itself and the normal component. If at this point the angular momentum stored in the superfluid component is released catastrophically, one has a sudden rise in the observed frequency of the pulsar, a `glitch' \citep{AI}.

Although the nature of the trigger mechanism for such glitches is not well understood, with hydrodynamical instabilities \citep{MMglitch, Nilsglitch}, crust-quakes \citep{Rud69} and vortex avalanches \citep{Alpartrigger, Lila11} all being candidates, the long timescales observed in the post-glitch relaxation in many systems provide strong evidence for a weakly coupled superfluid component \citep{2fluid} [see \citet{reviewHM} for a recent review of glitch models]. Furthermore recent studies of glitch statistics in radio pulsars have revealed that most follow a power-law distribution for the sizes and an exponential for the waiting times, with only three systems showing a degree of quasi-periodicity in the waiting times, and a preferred size for the glitches \citep{Melatos08} [although note that the small number of events preclude strong statistical conclusions, and it has been claimed that the distribution of glitch sizes in the Crab pulsar deviates from a power-law at the lower end \citep{CrabSize}]. Power-law size and exponential waiting time distributions are typical of Self Organised Critical (SOC) systems, wherein stresses built up over time by a global driver relax intermittently via nearest neighbour interactions \citep{Bak1}. \citet{lila08, lilaknock}, elaborating upon earlier suggestions by \citet{Alpartrigger} and \citet{capacitor}, have proposed that if vortices are `pinned' to lattice sites in the NS crust until a critical lag has built up, a SOC state can ensue. In this state the lag is close to the critical unpinning lag and, as a consequence of random fluctuations, vortices can unpin and knock on neighbouring vortices, causing avalanches and `glitches'. This suggestion has been borne out by Gross-Pitaevskii simulations of pinned superfluid vortices in a decelerating trap \citep{Lila11}.

The results of \citet{Lila11} provide support for the idea that vortices in the NS crust are in a SOC state. However the simulations are carried out in a regime that describes a Bose-Einstein condensate under laboratory conditions, but is not ideal for NS conditions. Such a setup is unavoidable due to the numerical difficulties associated with carrying out a full scale NS simulation. Several differences exist between the systems, ranging from the strength and range of the interaction, which are not appropriate for describing a strongly interacting system such as neutrons at high density, to the size of the simulation and model for the pinning potential.

One of the main issues in bridging the gap between the laboratory and a NS is the relative distance between vortices and pinning sites. In most simulations performed to date the inter vortex spacing is at most $\approx 10$ times the distance between pinning sites. In a NS, however, this is not the case. Pinning sites are separated by  $10^{-11}-10^{-10}$ cm, while vortices are spaced by roughly $10^{-3}$ cm for a radio pulsar spinning at $10$ Hz. It is thus natural to ask whether a vortex that unpins in a realistic NS setting would re-pin long before encountering another vortex, thus rendering `knock-on' effects \citep{lilaknock} irrelevant for a real star. If so, vortices will gradually move out, or `creep', hopping between adjacent pinning sites \citep{Alpar84a}, and vortex avalanches due to knock-on would not be viable as a trigger mechanism for pulsar glitches.

It is thus imperative to understand these issues if we are to accurately describe vortex motion in a realistic neutron star background and understand if the different manifestations (and different statistics) of glitching pulsars all stem from one underlying mechanism. In this paper we study in detail the hopping motion of pinned vortices, and  calculate the mean free path of a vortex in a 2-dimensional pinning potential, following the approach of \citet{Sedrakian95}. We also discuss the effects of vortex tension and bending, as 3-dimensional simulations by \citet{Link09} have shown that the self-energy of a vortex can have a strong impact on pinning; see also \citet{Tsubota} for a review on the role of vortex tension and turbulence in laboratory superfluids. In section \ref{s1} we present the equations of motion for a vortex and discuss unpinning and the effects of curvature in section \ref{unpinsec}. In section \ref{s3} we discuss vortex-vortex interactions, and the mean free path of a vortex in the crust is calculated in section \ref{s4}.

\section{Vortex motion}

\label{s1}
\subsection{Zero tension}
We begin by writing the equations of motion for a single vortex. Mathematically one can consider a superfluid vortex as a massless object and simply write an equation for force balance (see \citet{CB1, CB2} for a description of how to include an effective mass). It is important to note though that this is only a mathematical construction and that a vortex is not a material particle. The forces we write below are not localised; they are average forces acting on a suitably small volume around the vortex. With this in mind, the forces acting on a vortex will be the Magnus force,
\be
f_i^M=\kappa n_v \epsilon_{ijk} \hat{\Omega}^j(v^k_\mathrm{n}-v^k_\LL)\label{Magno};
\ee
a drag force,  due e.g. to the dissipative part of the interaction of the with the pinning sites,
\be
f_i^D=\kappa n_v \mathcal{R}(v^\mathrm{p}_i-v_i^\LL);
\ee
and the non-dissipative part of the interaction with the pinning site, i.e. the `pinning' force ${F}_i$. The equation of motion for a vortex thus takes the form:
\be
\epsilon^{ijk}\hat{k}_j (v_k^{\LL}-v_k^\n)+\mathcal{R}(v_\cc^i-v_\LL^i)+\mathcal{F}^i+\sigma_i=0,\label{gen}
\ee
where $v_\LL^i$ is the velocity of the vortex line, $v^i_\mathrm{p}$ is the velocity of the `normal' crust fluid of protons and electrons, $v^i_\mathrm{n}$ is the velocity of the superfluid neutrons,  $\rho_\n$ is the superfluid neutron mass density, $\mathcal{R}$ is the dimensionless drag parameter and we have $\mathcal{F}^i=F^i/{\rho_\n\kappa}$, with $\kappa$ the quantum of circulation and $\hat{\kappa}^i$ the unit vector along the vorticity , taken to be the $z$ axis in the following. The term $\sigma_i$ represents the `elastic' contribution of the vortex array (see e.g. \citet{Haskell11}), which derives from the fact that the local velocity field around the vortex will also depend on the positions of neighbouring vortices with respect to their equilibrium positions. We shall not consider $\sigma_i$ in this first investigation of vortex motion past a pinning site, but shall discuss how it can be approximated in section \ref{s3}. 

In the core mutual friction is mainly due to electrons scattering off vortex cores magnetised by entrainment \citep{AlparMF1}, which leads to a drag parameter of the form \citep{TrevMF}:
\be
\mathcal{R}\approx 4\times 10^{-4} \left(\frac{\delta m^*_\mathrm{p}}{m_\mathrm{p}}\right)^2\left(\frac{m_\mathrm{p}}{m^*_\mathrm{p}}\right)^{1/2}\left(\frac{x_\mathrm{p}}{0.05}\right)^{7/6}\left(\frac{\rho}{10^{14}\mbox{g/cm$^3$}}\right)^{1/6},\label{mfcore}
\ee
where $m^*_\mathrm{p}=m_\mathrm{p}(1-\varepsilon_\mathrm{p})$ is the effective proton mass, one has $\delta m^*_\mathrm{p}=m_\mathrm{p}-m^*_{\mathrm{p}}$, and $x_\mathrm{p}=\rho_\mathrm{p}/\rho$ is the proton fraction. In the crust, where electron scattering does not act, as protons are not superconducting, dissipation proceeds mainly via phonon interactions \citep{PhononJones} or, for larger vortex velocities, via Kelvin waves excited along the vortices themselves \citep{KelvonJones, KelvonBaym}.
The values of the crust's dissipation coefficients are, however, quite uncertain. They are likely to be low, $10^{-9} <\mathcal{R} <10^{-7}$ for phonon scattering, while kelvon processes are strongly dissipative, with $\mathcal{R} \approx 1$ \citep{Haskell12}.
If the protons in the outer core of the neutron star are in a type II superconducting state, additional damping can come from vortices cutting through flux tubes, a strongly dissipative process which leads to $\mathcal{R} \approx 10^{-3}$, although in this case the drag coefficient is actually dependent on the velocity of the vortices \citep{Link03, HaskellSat}. 

It is also important to note that for large velocities, vortices are likely to form a turbulent tangle, which will also change the form of the mutual friction force \citep{Peraltaglitch, melper07, TrevTurb}. An investigation of the problem in this case is beyond the scope of the current analysis. It should be the focus of future work, as it is likely to lead to quantitative and qualitative differences in the dynamics of the system. We also neglect the effect of the Bernoulli force due to the interaction of the superfluid flow around a vortex with ions in the crust. This interaction leads to additional repulsion between vortices and pinning sites, and will affect the estimates below for the velocity for a free vortex \citep{Alpar84a, BaymPin}. Although this effect is crucial for a correct description of vortex creep, it is, however, unlikely to strongly affect our results for re-pinning, which depend mainly on vortex motion inside the pinning potential.

From equation (\ref{gen}), the velocity of a vortex will be \citep{HaskellLagrange}:
\beq
v_\LL^i&=&v_\cc^i+\frac{1}{1+\mathcal{R}^2}(\mathcal{R}f^i+\epsilon^{ijk}\hat{k}_j f_k)\label{eos1}\\
f^i&=&\epsilon^{ijk}\hat{k}_j(v^\cc_k-v^n_k)+\mathcal{F}^i
\label{eos2}
\eeq

In the frame co-moving with the crust ($v^i_\cc=0$) this takes the form:
\beq
v_\LL^i&=&\frac{1}{1+\mathcal{R}^2}(\mathcal{R}f^i+\epsilon^{ijk}\hat{k}_j f_k)\label{eos3}\\
f^i&=&-\epsilon^{ijk}\hat{k}_jv^n_k+\mathcal{F}^i\label{eos4}
\eeq
We see immediately that a free straight vortex (i.e. outside the range of the potential) is forced to move mostly with the background neutron flow, with its velocity component along the flow  (i.e. along $\mathbf{v}^\n$) being:
\be
v^i_\parallel=(v_\LL^jv_j^\n)\hat{v}^i_\n=\frac{v^i_\n}{1+\mathcal{R}^2}\label{v1}
\ee
and the component perpendicular to the flow being:
\be
v^i_\perp=(\delta^i_j-\hat{v}_\n^i\hat{v}^\n_j)v^j_\LL
\ee
and
\be
{v}^i_\perp = -\mathcal{B}\epsilon^{ijk} \hat{k}_j v_k^\n \label{v2}
\ee
with
\be
\mathcal{B}=\frac{\mathcal{R}}{1+\mathcal{R}^2}
\ee

\subsection{Including tension}
\label{teso}
A vortex has a large self energy due to the kinetic energy of the flow around a curved segment of the vortex itself \citep{Thomson, Fetter}. Curvature introduces additional contributions to the neutron momentum, resulting in additional components of the neutron and proton velocities of the form \citep{TrevTurb}:
\beq
v_i^\n&=&\gamma_\n(\varepsilon_\n)\nu\epsilon_{ijk}\hat{k}^j\hat{k}^p\nabla_p\hat{k}^k\label{ten1}\\
v_i^\cc&=&-\gamma_\cc(\varepsilon_\n)\nu\epsilon_{ijk}\hat{k}^j\hat{k}^p\nabla_p\hat{k}^k\label{ten2}
\eeq
with 
\be
\nu=\frac{\kappa}{4\pi} \log{\left(\frac{a}{\xi}\right)},
\ee
where $\varepsilon_\n$ is the entrainment coefficient, $a$ is the inter vortex spacing, $\xi$ the coherence length associated with the vortex core, and one has \citep{TrevTurb}
\be
 \log{\left(\frac{a}{\xi}\right)}\approx 20 -\frac{1}{2} \log{\left(\frac{\Omega_\n}{100 \mbox{rad s$^{-1}$}}\right)}
 \ee
 which for the range of periods we are interested in (tens of milliseconds to seconds) is essentially a constant. We also define:
\beq
\gamma_\n&=&1-\frac{x_\cc\varepsilon_\n}{x_\cc-\varepsilon_\n}\\
\gamma_\cc&=&\frac{\varepsilon_\n(1-x_\cc)}{\varepsilon_\n-x_\cc}
\eeq
In the crust, where one expects small proton fractions ($x_\cc\ll 1$) and strong neutron entrainment [$\varepsilon_\n\approx 10$ \citep{Chamel1}] one has 
\beq
\gamma_\n&\approx& 1-x_\cc-\frac{x_\cc^2}{\varepsilon_\n}\\
\gamma_\cc&\approx&1-x_\cc+\frac{x_\cc}{\varepsilon_n}-\frac{x_\cc^2}{\varepsilon_\n}
\eeq
so that in the crust one has $\gamma_\n\approx\gamma_\cc\approx 1$. To account for vortex curvature and tension it is thus sufficient to add the contributions in (\ref{ten1}) and (\ref{ten2}) to the general flows in (\ref{eos1})-(\ref{eos4}).

\section{Vortex unpinning}

\label{unpinsec}

\subsection{Zero curvature}
Let us now focus on the special case of a vortex that is already pinned in the potential, and examine first of all the case of straight vortices. We consider a parabolic pinning potential $U_p$ with range $R_{range}$, and take the background superfluid flow to be along the $y$ axis, with $z$ the rotation axis. One then has:
\beq
U_p&=&\frac{1}{2}A(r_ir^i-r_0^ir_{0i})\;\;\;\mbox{for $|\mathbf{r}-\mathbf{r}_0|\leq R_{range}$}\\
U_p&=&0\;\;\;\mbox{for $|\mathbf{r}-\mathbf{r}_0|>R_{range}$}
\eeq
and the pinning force $F_p^i=-\nabla^i U_p$ takes the form
\beq
F_p^i&=&-A(r^i-r_0^i)\;\;\;\mbox{for $|\mathbf{r}-\mathbf{r}_0|\leq R_{range}$}\\
F_p^i&=&0\;\;\;\mbox{for $|\mathbf{r}-\mathbf{r}_0|>R_{range}$}\label{force}
\eeq
where $A$ is a constant that describes the strength of the interaction, $r^i$ is the position vector and $r_0^i$ is the position of the pinning site, which for simplicity we shall take at the origin in the following, i.e. $r_0^i=0$. Note that in the following we shall be dealing with a force {\it per unit length} acting on an (infinitely) long straight vortex, so that $U$ has units of g s$^{-2}$ cm$^{-1}$.
We are assuming the vortex is straight and infinitely long, so the general solution of equation (\ref{gen}) for the vortex position in the $x-y$ plane is:
\beq
x(t)\!\!&=&\!\!e^{-\mathcal{A}\mathcal{B}t}\left[C_1\cos\left(\frac{At}{\RR}\right)+C_2\sin\left(\frac{At}{\RR}\right)\right]+\nonumber\\
&&+\frac{V_y}{\mathcal{A}}\\
y(t)\!\!&=&\!\!e^{-\mathcal{A}\mathcal{B}t}\left[C_2\cos\left(\frac{At}{\RR}\right)-C_1\sin\left(\frac{At}{\RR}\right)\right]
\eeq
which are damped oscillations with a constant offset due to the Magnus force, with $C_1$ and $C_2$ constants. Note that $\mathcal{A}=A/(\rho_n\kappa)$.
Let us start by examining when a vortex originally at rest in the potential will unpin. 
In the absence of oscillations the condition is simply that the equilibrium position of the vortex be outside the potential range, i.e. that
\be
\left| \frac{V_y}{\mathcal{A}}\right|>R_{range}\label{unpin1}
\ee
Oscillations can, however, unpin a vortex before this threshold. Consider the case of a vortex originally at the centre of a potential, subject to an impulsive increase in the background flow from $v_\LL^y=0$ to $v_\LL^y=V_y$. In this case one has
\beq
x(t)=\!\!&=&\!\!-e^{-\mathcal{A}\mathcal{B}t}\left[\frac{V_y}{\mathcal{A}}\cos\left(\frac{At}{\RR}\right)\right]+\frac{V_y}{\mathcal{A}}\\
y(t)\!\!&=&\!\!e^{-\mathcal{A}\mathcal{B}t}\left[\frac{V_y}{\mathcal{A}}\sin\left(\frac{At}{\RR}\right)\right]
\eeq
If $\mathcal{R}\ll 1$ and the damping is weak, the distance from the centre of the potential, $R$, oscillates with time and takes the form:
\be
R\approx \sqrt{2}\left(\frac{V_y}{\mathcal{A}}\right)\sqrt{1-\cos\left({At}\right)}
\ee
so that oscillations can unpin the vortex (i.e. lead to $R>R_{range}$) for
\be
\left| \frac{V_y}{\mathcal{A}}\right|>\frac{R_{range}}{2}\label{unpin2}
\ee
We remind the reader once again that the result in (\ref{unpin2}) only holds if the damping timescale is longer than the period of the oscillations. The latter condition, $2\pi\mathcal{R}<<1$,  is generally satisfied for standard mutual friction mechanisms in the core [e.g. electron scattering off vortex cores \citep{AlparMF1}], but may not be satisfied if Kelvin waves are excited as vortices move past nuclear clusters \citep{KelvonJones}. We shall discuss this in more detail in the following section and derive a more general condition for unpinning in the presence of mutual friction.
Note that in the case of strong damping,  the condition for unpinning is simply $\left|{V_y}/{\mathcal{A}}\right|>R_{range}$, as in (\ref{unpin1}).

\subsection{Vortex curvature}
\label{curvo}
In the previous sections we have considered the motion of a straight vortex. It is however well known that tension plays an important role in vortex pinning, as a vortex can lower its energy by bending to intersect pinning bonds \citep{Link09}. The tension essentially sets the length-scale over which a vortex can be considered straight, with more rigid (i.e. longer) vortices leading to weaker pinning forces, as the difference in energy between different vortex/lattice configurations decreases \citep{JonesNOPIN, Stefano}. One can estimate the length scale over which a vortex remains straight by calculating when it is energetically favourable for a vortex to bend to pin to a nucleus. One thus has to compare the energy cost of bending over a certain distance $L$ to the energy gain of pinning to an additional nucleus at a distance $R_\mathrm{WS}$ (approximately the Wigner Seitz radius). This leads to \citep{HST, Stefano}:
\be
L\approx \frac{T_{se} R_\mathrm{WS}^2}{|E_p|}\approx 10^3 \;R_\mathrm{WS}\label{length}
\ee
where $E_p\approx 1$ MeV is the pinning energy and the tension $T_{se}=\rho_\n \kappa\nu \log{\left({a}/{\xi}\right)} \approx 20$ MeV/fm in the higher density regions ($\rho\approx 10^{14}$ g/cm$^3$) of the crust \citep{JonesNOPIN, TrevTurb}. In the lower density regions ($\rho\approx 10^{12}$ g/cm$^3$) one has $T_{se}\approx 1$ MeV/fm, and vortices remain rigid on shorter length scales. This should not be confused with the average distance between physical pinning sites obtained by \citet{LinkCutler}, but should rather be compared to the length scale over which a vortex unpins in the formulation of \citet{Link13}.

A hydrodynamical analysis confirms the estimate in (\ref{length}). We show in Appendix \ref{app1} that a vortex can bend to unpin over a length-scale
\be
L\gtrsim\sqrt{\frac{\nu R_{range}}{V_{cr}}}
\ee
which, if we assume that $R_{range}\approx R_\mathrm{WS}$ and approximate the critical velocity as $V_y^\n\approx E_p/(\rho_\n\kappa R_\mathrm{WS} L)$, leads to $L\approx (T_{se} R^2_\mathrm{WS})/E_p\approx 10^3 R_\mathrm{WS}$ in the deep crust as estimated from energetics in (\ref{length}).

In practice we expect the system to self-adjust to always be close to the critical unpinning threshold, and small variations in pinning strength and thermal unpinning rates are likely to influence the unpinning process \citep{Lila09}.

The general picture we adopt is thus that a vortex that unpins over a length scale $L\approx T R^2_\mathrm{WS}/E_p$, and subsequently moves freely with the background flow, unzipping as it moves out, until it repins.  The main effect of the tension is to set the length-scale over which the vortex unpins and effectively renormalise the strength of the pinning interaction \citep{Link09, Stefano}. Luckily we will see that our results depend only weakly on the exact value of the pinning force. When required we will use the values calculated by \citet{Stefano} who account for the reduction of the pinning force due to the finite rigidity of vortices.

\section{Vortex lattice forces}
\label{s3}

In a neutron superfluid vortices (in the absence of pinning) are expected to form a triangular lattice that maximises the distance between each vortex, thus minimising vortex-vortex interactions due to the irrotational flow around a vortex core. In the presence of pinning the distribution of vortices may deviate from such a lattice, but we will assume that the high relative abundance of pinning sites with respect to vortices always allows for the creation of a triangular vortex lattice.

In their equilibrium positions in the lattice vortices are exposed to the average background flow generated by all other vortices. Displacements from the equilibrium position will move a vortex closer to one of its neighbours, and there will be an additional local contribution to the velocity field from the irrotational flow around its axis.

For small displacements from the equilibrium lattice position we can regard the vortices as connected by springs, such that the force per unit length between acting on a vortex $a$ is:
\be
\mathbf{F}_e={K}\sum_{a\neq b}(\mathbf{r}^{ba}-\mathbf{r}^{0,ba})
\ee
with $\mathbf{r}_{ba}=\mathbf{r}_b-\mathbf{r}_a$ and $a$, $b$ label the vortices, $\mathbf{r}^{0,ab}$ the equilibrium separation between them (which we take to be the Feynman distance) and the elastic constant $K$ such that $K=4\mu/\sqrt{3}$ \citep{Lee1990}, where $\mu$ is the shear modulus of a vortex array \citep{CB1}:
\be
\mu=\frac{\rho_\n\kappa\Omega}{16\pi}.
\ee
The term $\sigma_i$ in (\ref{gen}) thus takes the form:
\be
\sigma_i=\frac{4\Omega}{16\sqrt{3}\pi}\sum_{a\neq b}(r^{ba}_i-r^{0,ba}_i)
\ee
This form of the interaction is appropriate for the description of small oscillations around the equilibrium position of a vortex in the lattice, `Tkachenko' waves, which may have observable consequences in radio pulsars \citep{Haskell11}.

\begin{figure}
\centerline{\includegraphics[scale=0.8]{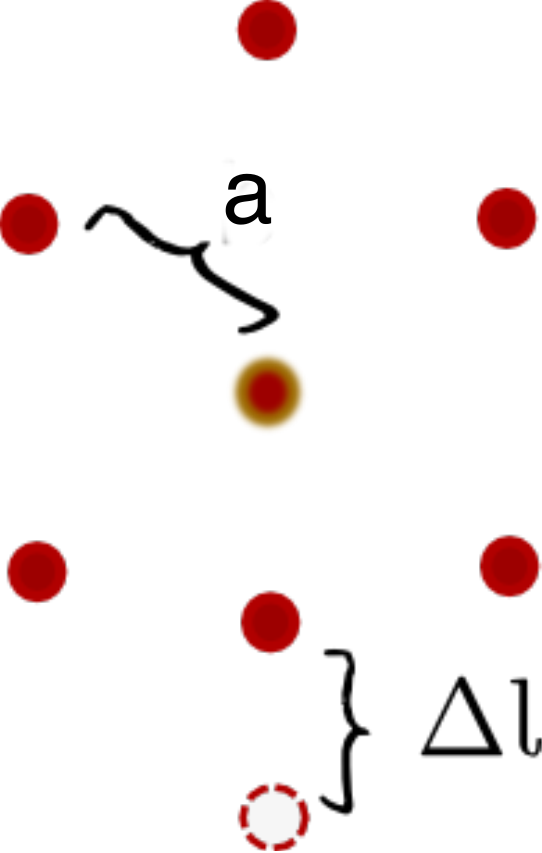}}
\caption{A schematic representation of the displacement of a vortex from its equilibrium position in an Abrikosov lattice. The equilibrium inter-vortex spacing is $a$ and the vortex is displaced a distance $\Delta l$, thus moving closer to other vortices and perturbing the Magnus force, as described in the text.}\label{vortexL1}
\end{figure}

If, however, a vortex can unpin and move a distance comparable to the inter-vortex spacing $a$, the `elastic' description breaks down. In this case we can estimate the contribution of the nearest neighbours to the force acting on a vortex by considering a simple hexagonal cell in which the vortex is displaced in the y direction (aligned with the background flow) by an amount $\Delta l$, as in figure \ref{vortexL1}. We remind the reader that we are assuming that pinning sites are abundant enough to allow for vortices to organise in a triangular lattice without disruption.

If a vortex is displaced from its equilibrium position there will be a perturbation in the superfluid velocity field and thus in the Magnus force acting on all other vortices. For simplicity in the following we shall analyse the velocity perturbation and compare it to the background flow. Given that the Magnus force scales linearly with the velocity of the superfluid flow past a vortex (see eq. \ref{Magno}), this is equivalent to studying when the additional contribution to the Magnus force acting on a vortex becomes comparable to the Magnus force due to the background flow. For $\Delta l/a <1$ the additional contribution $\Delta \mathbf{v}$ to the background superfluid velocity  ${V}_i^\n$ from the displacement is simply:
\be
|\Delta \mathbf{v}| \approx \frac{\kappa}{2\pi a} \frac{\Delta l}{a}\approx  0.1 \left(\frac{f^r}{10\mbox{Hz}}\right)^{-1/2} \left(\frac{\Delta l}{a_f}\right) \mbox{cm/s}\label{vario}
\ee
where we have taken $a_f\approx 4\times 10^{-3}/\sqrt{(f^r/10\;\mbox{Hz})}$ cm, with $f^r$ the rotation frequency in Hz.
In order to understand the relevance of this perturbation to the dynamics of vortices in the crust we must compare it to typical critical velocities for unpinning, which are in the range $V_{cr}\approx 10^2-10^4$ cm/s \citep{SPH12}. As the system self adjusts to remain close to the unpinning threshold, we can expect the average condensate velocity to settle close to the critical unpinning  velocity, with $(V_{cr}-{V}^\n)/V_{cr}\approx 0.001-0.01$, depending on the temperature of the star and the local pinning strength \citep{Alpar84a}.
In conclusion the perturbation to the Magnus force acting on a vortex could be sizeable compared to the background value, given that the system will be close to the critical unpinning threshold, and we expect:
\be
10^{-5} \lesssim \left|\frac{\Delta v}{V_{cr}-V^\n}\right|\lesssim 1 \label{vario}
\ee

Given the results above, in the following we will thus use $\Delta l/ a\approx 1$ as our criterion to decide whether avalanches can propagate or not. Nevertheless one should keep in mind that, as we can see from equation (\ref{vario}), vortex avalanches could also propagate for $\Delta l \ll a$ if the system is very close to the critical unpinning threshold  \citep{lilaknock, Lila12}.

\begin{figure}
\centerline{\includegraphics[scale=0.8]{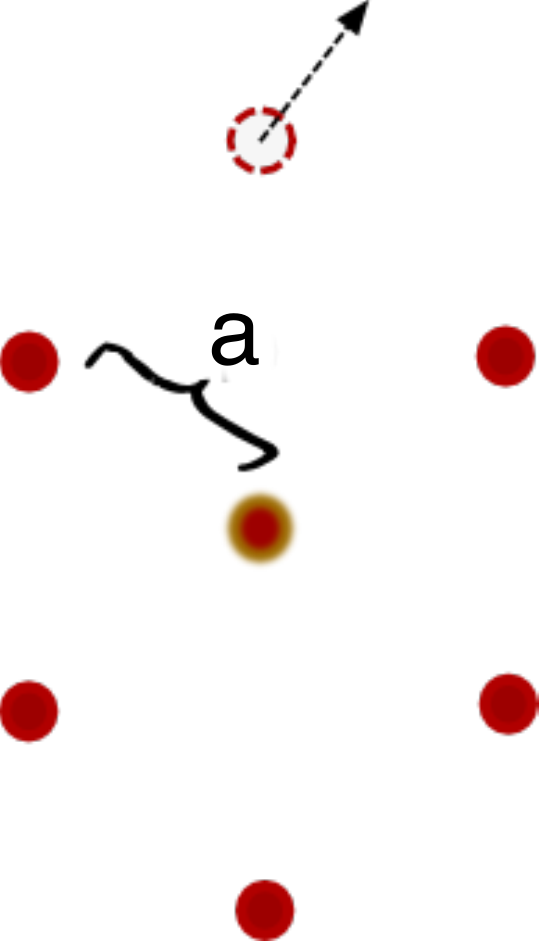}}
\caption{A schematic representation of the `suction' effect, in which a vortex has escaped it's equilibrium position in an Abrikosov lattice, with equilibrium inter-vortex spacing $a$. This perturbs the Magnus force acting on the remaining vortices, possibly unpinning them and leading to a backward (i.e. uphill with respect to the flow) propagating avalanche seeking to fill the hole.}\label{vortexL2}
\end{figure} 
In the previous discussion we have considered a vortex that moves a distance $\Delta l$ and unpins a neighbouring vortex, leading to a knock on effect. It is important to note, however, that together with this forward propagating avalanche, one can have a backward propagating avalanche due to the `suction' effect.
Consider the situation in figure \ref{vortexL2} in which a vortex has broken free and moved away, leaving a gap in the array. The contributions from the Magnus force acting on the central vortex are no longer balanced, leading to a net velocity perturbation
\be
\Delta v\approx  \frac{\kappa}{a}\approx - 0.1 \left(\frac{\nu}{10\mbox{ Hz}}\right)^{-1/2} \mbox{cm/s}\label{vario2}
\ee
which, as discussed before, can be significant compared to the background flow. In fact Gross Pitaevskii and N-body simulations of vortex motion reveal that  backward propagating avalanches due to the suction effect are key drivers of vortex unpinning, also for $\Delta l < a$ \citep{JamesGW}.

\section{Vortex re-pinning}
\label{s4}

We have seen in section \ref{curvo} that a vortex unpins over a length scale $L$, and then, if the system is close to the unpinning threshold (as one would expect from a self-organised critical system), it  moves essentially freely with the flow, with tension renormalising the unpinning threshold.  The main question to address now is whether the vortex will re-pin immediately, or whether it is generally free to move past enough pinning sites that it can get close enough to other vortices to trigger an avalanche. 

Let us simplify the problem by considering a free, straight, vortex in the $x-y$ plane, entering the range of a pinning potential. It is very important to note that this 2-dimensional potential describes the change in energy of adjacent configurations of a rigid vortex, which will differ by the number of intersected nuclei, and is {\bf not} the pinning potential due to individual pinning sites. In fact a rigid vortex which is straight on a length scale $L\ll R_{WS}$ will always intersect a number of nuclei as it moves, and pinned configurations will be those in which more nuclei are intersected \citep{Stefano}. The potential will thus depend on the shape of the crystal lattice, but also crucially on vortex orientation and rigidity, and is thus likely to resemble a random potential even in the presence of a regular lattice.

For such a potential quantum tunnelling to adjacent configurations is disfavoured, as the probability decreases exponentially as the distance between pinned configurations increases, and is small compared to the probability of thermal unpinning at a few coherence radii \citep{Auer06, Fialko12}. For vortices that can be considered `rigid' on length scales $L\approx 100-1000 R_\mathrm{WS}$, adjacent configurations are separated by approximately 10-100 coherence lengths \citep{Stefano}, so we shall not consider the small probability of tunnelling [see \citet{Link93} for an in-depth discussion], but rather consider the vortex as a classical object as in \citet{Link13}, and calculate its re-pinning cross section.

\begin{figure*}
\centerline{\includegraphics[scale=0.43]{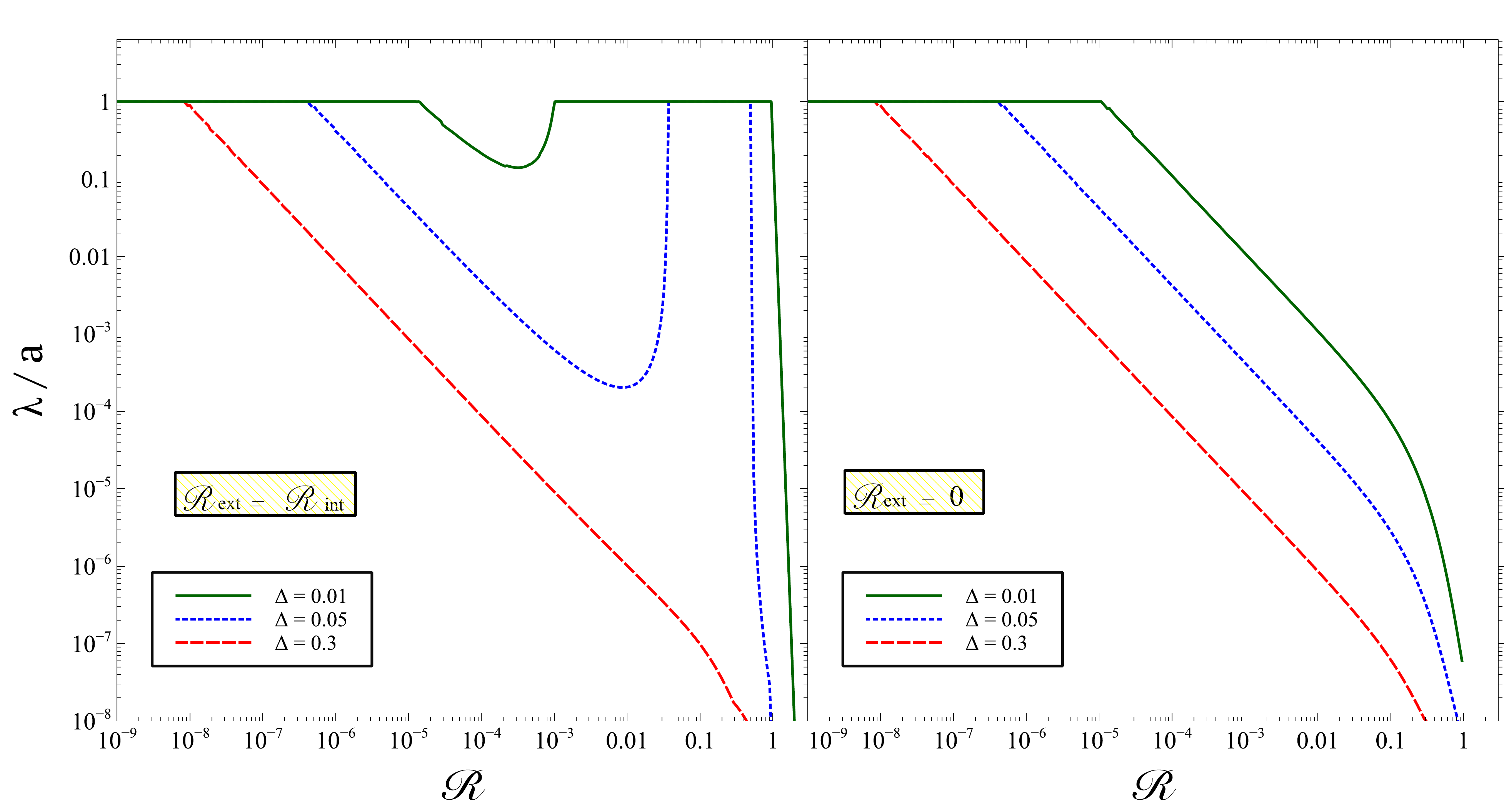}}
\caption{Mean free path $\lambda$ (normalised to the inter-vortex spacing $a$) of a vortex before re-pinning, for varying mutual friction parameter $\mathcal{R}$, and for different values of $\Delta=(V_C-V^\n_y)/V_C$.  In the left panel the strength of the mutual friction is the same both inside and outside of the range of the potential, while in the right panel we have assumed that there is no damping outside the range of the potential, for $r>R_{range}$. We can see that if mutual friction is the same everywhere (left panel) avalanches can propagate for $\Delta\lesssim 0.05$ for most values of $\mathcal{R}\lesssim 1$. If damping is weak outside the range of the potential (right panel) avalanches can only propagate for weak values of the mutual friction parameter. This somewhat counter-intuitive result is due to the fact that mutual friction also restricts the range of impact angles, as described in the text.}\label{lambda}
\end{figure*} 

\subsection{Single-vortex motion}

Let us consider a vortex entering the range of the potential at a position $(R_{0}\cos\theta, R_{0}\sin\theta)$ in the plane. The solution for its subsequent motion is
\beq
x(t)=\!\!&=&\!\!e^{-\mathcal{A}\mathcal{B}t}\left[\left(-\frac{V_y}{\mathcal{A}}+R_{0}\cos\theta\right)\cos\left(\frac{At}{\RR}\right)\right.\nonumber\\
&&\left.+R_{0}\sin\theta\sin\left(\frac{At}{\RR}\right)\right]+\frac{V_y}{\mathcal{A}}\label{repin1}\\
y(t)\!\!&=&\!\!e^{-\mathcal{A}\mathcal{B}t}\left[R_{0}\sin\theta\cos\left(\frac{At}{\RR}\right)\right.\nonumber\\
&&-\left.\left(-\frac{V_y}{\mathcal{A}}+R_0\cos\theta\right)\sin\left(\frac{At}{\RR}\right)\right]\label{repin}
\eeq

The first thing to note is that if the condition in (\ref{unpin1}) is satisfied the vortex will necessarily unpin again (i.e. return outside the range of the pinning potential) after oscillating in the potential, regardless of the damping strength $\mathcal{R}$. This is to be expected as, given its vanishing inertia, the motion of the vortex is entirely determined by its position (it instantaneously adapts to force balance). Mathematically this can be seen in the fact that we are solving a first order differential equation, rather than a second order one, so there is only one constant to set (i.e. the initial position), as opposed to a standard scattering problem which has two constants (e.g. the impact parameter and the velocity of the particle).

\begin{figure}
\centerline{\includegraphics[scale=0.6]{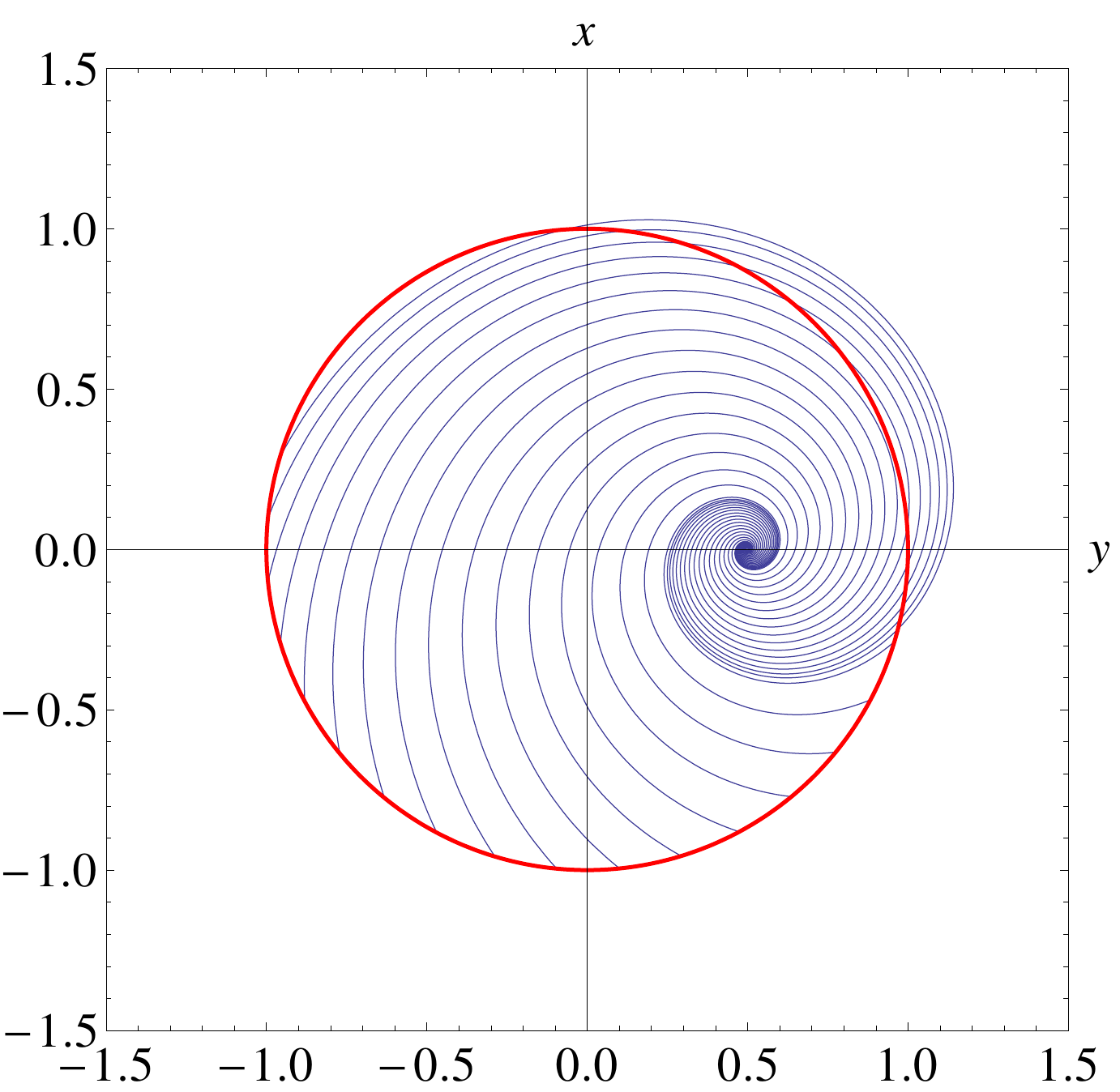}}
\caption{Trajectory of a vortex initially at the edge of the potential ($x^2+y^2=1$) for varying impact angle. We have taken $R_{range}=1$. We consider relatively strong damping with $\mathcal{R}=0.3$ and a background flow along the $y$ axes, with $V_y^\n/\mathcal{A}=R_{range}/2$. We see that for our choice of reasonably strong mutual friction there is a range of impact angles ($\theta>5\pi/4$) in the lower quadrant for which a vortex cannot make it back out of the potential and is forced to re-pin, eventually settling down at its equilibrium position at $x=R_{range}/2$.}
\label{traj}
\end{figure}

However, even if the system is below the unpinning threshold in (\ref{unpin1}), in the absence of damping a vortex will always oscillate out of the pinning potential with exit angle $\phi_{ND}=2\pi-\theta$ after a time $t_{ND}=2\alpha/\omega$, where 
\be
\tan \alpha=\frac{\sin\theta}{\cos\theta-V^\n_y/(\mathcal{A} R_{range})}
\ee
and
\be
\omega=\frac{\mathcal{A}}{\RR}
\ee
Even for very small values of the background flow there will be no repinning, which is essentially the conclusion of \citet{Sedrakian95}, who also studied re-pinning of straight vortices in 2D.

In the presence of weak damping, i.e. $\mathcal{R}\ll 1$, we can approximate the exit angle $\phi$ and exit time $t$ as $\phi=\phi_{ND}+\delta\phi$ and $t=t_{ND}+\delta t$, which leads to:
\beq
\delta\phi&=&{2\alpha\mathcal{R}}\frac{\mathcal{A} R_{range}}{\sin\theta V_y^\n}\left[ \left(\cos\theta-\frac{V_y^\n}{\mathcal{A} R_{range}}\right)^2+\sin^2\theta \right]\label{delta}\\
\delta t&=&\frac{2\alpha\mathcal{R}}{\omega}\left[\frac{\sin\alpha\sin(\alpha-\theta)-1}{\cos\alpha\sin(\alpha-\theta)}\right]\label{delta2}
\eeq

There is a range of impact angles close to $\theta=0$ for which escape is no longer possible, an example of which can be seen by the plot of the trajectories in figure \ref{traj} for $\mathcal{R}=0.3$. The range of impact angles for which the vortex re-pins  depends strongly on the mutual friction parameter $\mathcal{R}$ through (\ref{delta}) and (\ref{delta2}), as does the range of impact angles that are possible, as a free vortex between pinning sites travels at an angle $\theta_d=\tan^{-1}( \mathcal{R})$ with the $y$ axis, as can be seen from equations \ref{v1} and \ref{v2}.

Two scenarios are possible: either the mutual friction strength  $\mathcal{R}$ is the same everywhere, or else the value of $\mathcal{R}$ for $r<R_{range}$ (i.e. inside the range of the pinning potential) is much stronger, as Kelvin waves are excited as the vortex moves past the nuclei and bends to adapt to a pinned configuration \citep{Link09}  [although it is also possible that in this case vortices will also form a tangle \citep{melper07, TrevTurb}]. Note that this is only the {\it local} drag acting on the vortex. The overall value of the Mutual Friction parameter $\mathcal{R}$, on a macroscopic hydrodynamical scale, depends also on the number of free vortices one averages over to obtain a hydrodynamical description of the superfluid, i.e. there is an effective drag $\tilde{\mathcal{R}}=\gamma\mathcal{R}$, with $\gamma$ the fraction of vortices that are free. If most vortices are pinned  the overall drag can be quite low ($\gamma \ll 1$), leading to long coupling timescales for the superfluid and the crust, even though locally a vortex experiences strong damping \citep{JM06, Haskell12}.

\subsection {Effective cross-section}

We compute the re-pinning cross section $\sigma$ by numerically solving equations (\ref{repin}) for the range of impact angles $\theta$ for which re-pinning is possible. We then compute the mean free path of a vortex, calculated as
\be
\lambda_v=\frac{1}{n_p \sigma}
\ee
where $n_p$ is the number of pinning sites per unit volume, assumed to be cylindrically symmetric and regularly spaced by $2R_{range}$. Unless otherwise stated, we assume that $R_{range}=100$ fm. 
This approach is justified if successive encounters between a pinning site and a vortex are uncorrelated. In a periodic pinning lattice correlations arise; in Appendix \ref{app2} we show that in this case vortices span the whole range of entry and exit angles as they `hop' between pinning sites. The pinning potential in a neutron star is likely to resemble a random potential, both due to physical irregularities in the crust \citep{JonesAm, Kp14} and due to the random orientations of the vortex with respect to the lattice \citep{Stefano}. In this case long range correlations are less likely.

The results depend on three parameters: the impact angle $\theta$ (whose range depends on the mutual friction strength outside the pinning potential), the ratio $V^\n_y/\mathcal{A}R_{range}$, which quantifies how close the system is to the critical unpinning velocity, and  $\mathcal{R}$ the strength of the mutual friction inside the pinning potential.

How close the global hydrodynamical system can come to the unpinning threshold will depend on the vortex motion itself, and in general can only be determined by running large scale Gross-Pitaevskii simulations such as those in \citet{Lila11}. In the following we will thus simply treat $V^\n_y/\mathcal{A}R_{range}$ as a parameter. In order to obtain an estimate of this parameter we can, however, estimate how efficient vortex creep is at removing vorticity and locking the two fluids together. At high temperature many vortices will unpin and `creep' out, leading to a high fraction $\gamma$ of vortices that are free on average (although note that the majority of vortices will still remain pinned). This could give rise to a strong enough effective drag $\tilde{\mathcal{R}}=\gamma\mathcal{R}$ to allow the superfluid to spin down together with the crust. The overall lag of the fluid region will thus be fixed well below the critical unpinning threshold and we may expect vortices to trickle out gradually. If, on the other hand, creep is inefficient most vortices are pinned and the effective drag will be low. It cannot stop the fluid from getting globally very close to the global threshold for unpinning and then the superfluid will expel vorticity by means of discrete `avalanche' events \citep{lilaknock}. Essentially there are two ways that the superfluid can follow the spin-down of the crust: either a high unpinning rate (due for example to thermal fluctuations) leads to vortices gradually moving out via many small events (vortex creep), or one has a low unpinning rate and large avalanches. The spin down rate $\dot{\Omega}_\n$ of a pinned superfluid coupled to the normal fluid can be approximated as:
\be
\dot{\Omega}_\n\approx\gamma 2\Omega_\n \frac{\mathcal{R}}{1+\mathcal{R}^2} \Delta\Omega_{cr}.\label{lock1}
\ee
where $\Delta\Omega_{cr}$ is the critical lag $\Omega_\cc-\Omega_\n$ for unpinning and
\be
 \gamma\approx \exp\left(-\frac{E_a(\Delta\Omega_{cr}-\Delta\Omega)/ \Delta\Omega_{cr}}{kT}\right)
 \ee

$T$ is the temperature in the inner crust and $E_a$ is the activation energy for unpinning \citep{LinkUnpin}, which in the following we take to be approximately the pinning energy, so that $E_a=E_p\approx 1$ MeV. In order for the time-average lag to be stationary one must have:
 \be
 \dot{\Omega}_\n=\dot{\Omega}_\cc={\dot{\Omega}}_O\label{lock2}
 \ee
 where $\dot{\Omega}_O$ is the observed spin-down rate. For typical parameters of glitching pulsars (${\dot{\Omega}}_O=-10^{-10}$ s$^{-2}$, $\Omega=100$ s$^{-1}$, $E_p=1$ MeV, $\Delta\Omega_{cr}=10^{-3}$ rad/s) one has that for hotter stars ($T=10^8$ K) already for $(\Delta\Omega_{cr}-\Delta\Omega)/ \Delta\Omega_{cr})<0.01$ the creep rate is high, i.e. $\gamma\approx 1$, and the condition on the drag for the two fluids to be coupled can be obtain be combining equations (\ref{lock1}) and (\ref{lock2}), and is $\mathcal{R}\gtrsim 10^{-9}$. The values of $\mathcal{R}$ in the crust are highly uncertain, but as can be seen from the estimates in table \ref{nv}, this condition should be satisfied in lower density regions of the crust. For colder pulsars ($T=10^7$ K), on the other hand, even for  $(\Delta\Omega_{cr}-\Delta\Omega)/ \Delta\Omega_{cr})\approx 0.01$ the creep rate is low and one has the condition $\mathcal{R}\gtrsim10^{-5}$ which may not be satisfied if the main mutual friction mechanism in the crust, for slow vortex motion, is coupling to phonon excitations of the lattice. Creep will then be ineffective and the lag will increase to $(\Delta\Omega_{cr}-\Delta\Omega)/ \Delta\Omega_{cr}\approx 10^{-3}$ before one has $\gamma\approx 1$ and creep can become effective again. 
However, even if $\gamma\approx 1$ the lower estimates of the mutual friction strength in table \ref{nv} suggests that there may be regions in the crust in which creep cannot lead to a strong enough coupling and the superfluid always needs to expel vorticity via discrete avalanches. A better understanding of mutual friction mechanisms in the crust is thus imperative in order to understand to what extent vortex-vortex interactions can affect the dynamics of the star.

Let us stress again, however, that those presented above are simply estimates and our approach {\bf cannot} determine $(\Delta\Omega_{cr}-\Delta\Omega)/ \Delta\Omega_{cr})$, which in the following will we treat  simply as a parameter. We calculate the cross section for the case in which the background flow is $30\%$ less than the global critical velocity in (\ref{unpin1}), and also examine the cases in which the background flow it is $5\%$ , $1\%$ and $0.1\%$ less than the critical velocity.

\begin{table}
\begin{tabular}{lllllllll}
\hline
Region&1&2&3&4&5\\
\hline
&&&&&\\
$\rho/(10^{14}\mbox{g/cm$^3$})$&0.015& 0.096& 0.34&0 .78& 1.3\\
&&&&&\\
$\mathcal{R}_K$&2.9&0.058&3.4&0.00085&0\\
&&&&&\\
$\mathcal{B}_K$&0.31&0.06&0.3&0.00085&0\\
&&&&&\\
$\mathcal{B}_P/10^{-9}$&$1.6$&$0.7$&$5.8$&$0.025$&0\\
&&&&&\\
$R_{range} (\mbox{fm})$&167&131&97&88&69\\
\hline
\end{tabular}
\caption{Fiducial values of the density $\rho$, range of the pining potential $R_{range}$, Kelvon drag parameter $\mathcal{R}_K$, mutual friction parameter $\mathcal{B}_K=\mathcal{R}_K/(1+\mathcal{R}_K^2)$ and phonon mutual friction parameter $\mathcal{B}_P=\mathcal{R}_P/(1+\mathcal{R}_P^2)\approx \mathcal{R}_P$, for five regions of the crust from \citet{NegVau}  and \citet{SPH12, Stefano}}\label{nv}
\end{table}

The results of our calculations are shown in figure \ref{lambda}, where we plot the ratio $\lambda/a$ of the mean free path to the inter-vortex spacing as a function of $\mathcal{R}$. For ease of presentation, and given that it does not make a difference for our criterion for avalanches, we cutoff our results at $\lambda/a=1$, although the mean free path in some cases is much larger than the inter vortex separation. The first thing to note is that if $\mathcal{R}\gtrsim 1$ vortices repin immediately. This is mainly because the period of oscillations is proportional to $1+\mathcal{R}^2$, while the damping timescale is proportional to $1/\mathcal{B}=(1+\mathcal{R}^2)/\mathcal{R}$. For large values of $\mathcal{R}\gg 1$ the damping timescale is thus much shorter than the oscillation timescale, and the vortex does not have time to complete an oscillation around the pinning centre before it stops moving. 

In the case $\mathcal{R}\lesssim 1$ we see that there can be a significant difference between the two damping scenarios we consider. If the mutual friction is the same both outside and inside the pinning potential range, then there is a vast range of mutual friction strengths for which the mean free path is greater than the inter-vortex spacing, already if the background flow is $5\%$ less than the critical unpinning threshold. Once the difference is less than $1\%$ vortices are essentially always able to move close to each other and cause avalanches.
Things however change drastically if we assume that mutual friction is much weaker outside the pinning potential. From figure $\ref{lambda}$ we see that if $\mathcal{R}=0$ outside the pinning range then re-pinning becomes much more likely and avalanches are only possible for very weak drag parameters ($\mathcal{R}\lesssim 10^{-6}$) or if $(V_C-V^\n_y)/V_C\lesssim 0.001$.

\begin{figure}
\centerline{\includegraphics[scale=0.59]{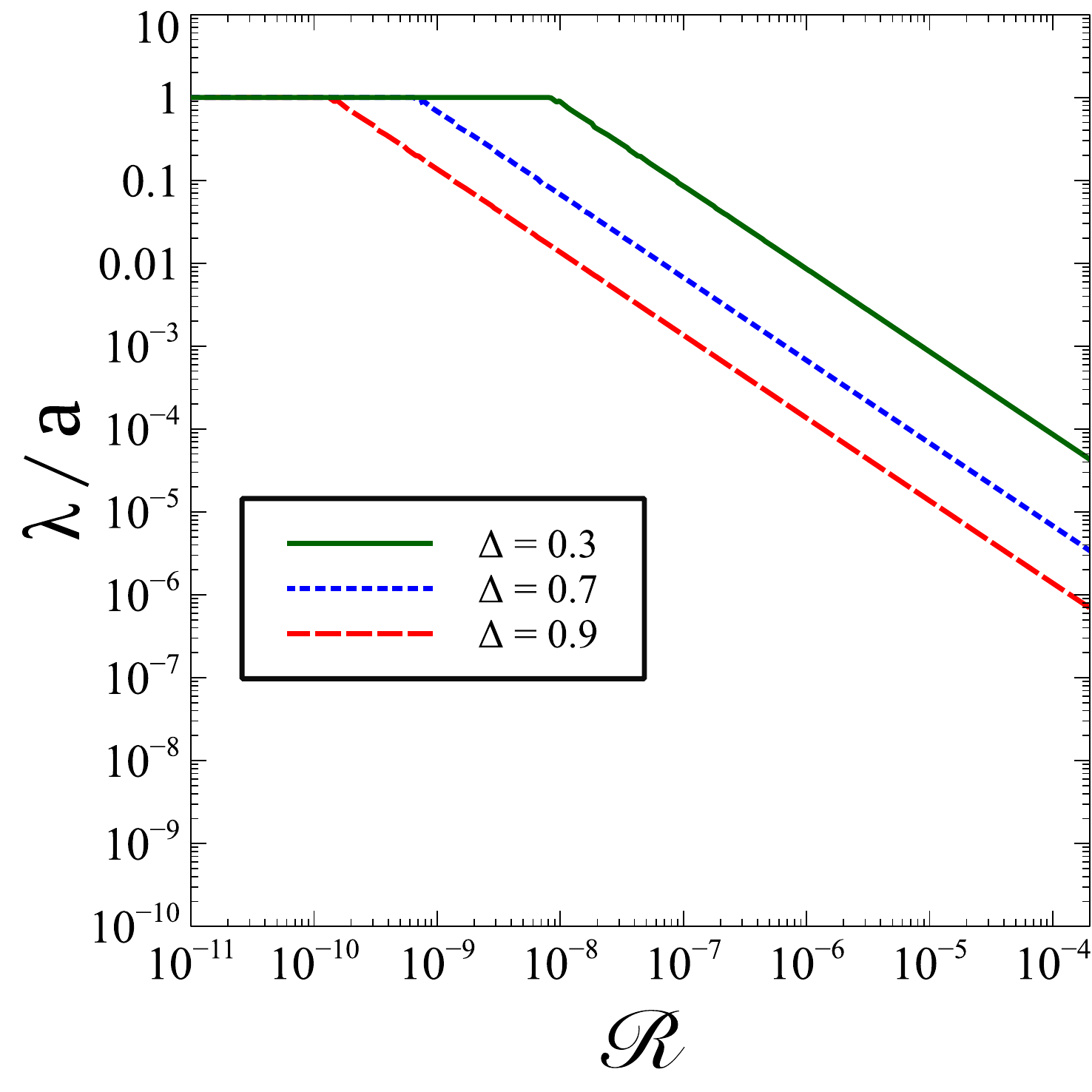}}
\caption{Mean free path $\lambda$ (normalised to the inter-vortex spacing $a$) of a vortex before repinning, for varying mutual friction parameter $\mathcal{R}$, and for different values of $\Delta=(V_C-V^\n_y)/V_C$. We see that for weak drag avalanches are possible even for large $\Delta$, i.e. even if the system is far from the mean unpinning threshold.}\label{weak}
\end{figure}

In the case of very weak damping ($\mathcal{R}\lesssim 10^{-8}$), in fact, one always has $\lambda\approx a$, even for unpinning at background neutron velocities $V_\n$ an order of magnitude lower than the global critical unpinning velocity, as can be seen in figure \ref{weak}. It is thus plausible that even vortices that unpin from weaker pinning sites well below the critical threshold could trigger an avalanche, if they are moving too slowly to excite Kelvin waves and are subject to weak drag.

The value of $\mathcal{R}$ in a NS crust is poorly known. However, to get a qualitative idea of what kind of dynamics to expect in a realistic neutron star we calculate the mean free path for the values of $\mathcal{R}$ due to Kelvon drag calculated by \citet{Haskell12}, following the prescription of \citet{KelvonJones}, which are given in table \ref{nv}.
As can be seen from tables \ref{restab1} and \ref{restab2} most vortices in the lower density regions of the crust will repin to adjacent pinning sites, due to the strong damping and the abundance of free sites. In the higher density regions, however, vortices can move a considerable fraction of the inter vortex spacing before repinning. In fact in region 5, the highest  density region in the table, the mean free path is formally the size of the system, due to the weak damping.

Depending on models for the pairing gap of superfluid neutrons used in calculations, the maximum of the pinning force acting on the vortices will occur between regions 3 and 4. Re-pinning in the inner regions (4 and 5)  will then be more likely than we estimate in tables \ref{restab1} and \ref{restab2}, as vortices moving out will encounter a generally increasing pinning potential as they approach the maximum of the pinning force \citep{Stefano, SPH12}, thus decreasing the mean free path. The potential, however, varies over a much longer length scale than the inter vortex spacing and the mean free path, so we ignore this effect in the current calculation.

\begin{table*}
\begin{tabular}{clllllllll}

& & & \multicolumn{4}{c} {$\boldsymbol{\lambda}\mathbf{/a}$}\\[3pt]
 \cline{4-7}\\
{\bf Region} & $\dfrac{\boldsymbol{\rho}}{\mathbf{(10^{14}}\mbox{{\bf g/cm}$\mathbf{^3}$}\mathbf{)}}$ && $\Delta=0.3$ & $\Delta=0.05$ & $\Delta=0.01$ & $\Delta=0.001$\\
\hline\\[2pt]
1 & 0.015 && - & - & - & $7.5\times 10^{-9}$\\[4pt]
2 & 0.096 && $1.7\times 10^{-7}$ & $8.8\times 10^{-6}$ & $2.3\times 10^{-4}$ & $2.3\times 10^{-2}$\\[4pt]
3 & 0.34 && - & - & - & $5.4\times 10^{-9}$\\[4pt]
4 & 0.78 && $1.0\times 10^{-5}$ & $5.0\times 10^{-4}$ & $1.3\times 10^{-2}$ & $1.3$\\[4pt]
5 &1.3 && $\ll 1$ & $\ll 1$ & $\ll 1$& $\ll 1$\\[4pt]

\end{tabular}
\caption{Mean free path $\lambda$, normalised to the inter-vortex spacing $a$, for the regions described in table \ref{nv} and external mutual friction parameter $\mathcal{R}_{ext}=0$. We see that in this case for all values of $\Delta=(V_C-V^\n_y)/V_C$ vortex avalanches are unlikely in all but the highest density regions, in which the mean free path is formally the size of the system due to weak pinning and drag.}
\label{restab1}
\end{table*}

\begin{table*}

\begin{tabular}{clllllllll}

& & & \multicolumn{4}{c} {$\boldsymbol{\lambda}\mathbf{/a}$}\\[3pt]
 \cline{4-7}\\
{\bf Region} & $\dfrac{\boldsymbol{\rho}}{\mathbf{(10^{14}}\mbox{{\bf g/cm}$\mathbf{^3}$}\mathbf{)}}$ && $\Delta=0.3$ & $\Delta=0.05$ & $\Delta=0.01$ & $\Delta=0.001$\\
\hline\\[2pt]
1 & 0.015 && - & - & - & $2.3\times 10^{-8}$\\[4pt]
2 & 0.096 && $3.4\times 10^{-7}$ & $>1$ & $>1$ & $>1$\\[4pt]
3 & 0.34 && - & - & - & $3.2\times 10^{-8}$\\[4pt]
4 & 0.78 && $1.1\times 10^{-5}$ & $7.0\times 10^{-4}$ & $0.42$ & $>1$\\[4pt]
5 &1.3 && $\ll 1$ & $\ll 1$ & $\ll 1$& $\ll 1$\\[4pt]

\end{tabular}
\caption{Mean free path $\lambda$, normalised to the inter-vortex spacing $a$, for the regions described in table \ref{nv} and for $\mathcal{R}_{ext}=\mathcal{R}_{int}$, i.e. equal mutual friction strength inside and outside the range of the potential. In this case avalanches appear possible in regions 2, 4 and 5 for values of  $\Delta=(V_C-V^\n_y)/V_C\lesssim 0.01$.}
\label{restab2}
\end{table*}

\section{Conclusions}

In this paper we have calculated the mean free path of a superfluid neutron vortex as it scatters off pinning sites in the crust of a neutron star, to assess the viability of vortex avalanches as a trigger mechanism for pulsar glitches. This is a crucial point as Gross-Pitaevskii simulations of a pinned Bose-Einstein condensate in a decelerating trap show that the system settles down in a self organised critical state, in which the superfluid spins down by a series of discrete vortex avalanches \citep{lilaknock}. This paradigm naturally explains why the distribution of glitch sizes in most pulsars is well approximated by a power-law \citep{Melatos08}. However, due to computational limitations, vortices are separated by tens of pinning sites in a typical Gross-Pitaevskii simulation,  compared to $10^{10}$ pinning sites in a realistic neutron star crust. It is thus natural to ask whether a vortex avalanche would be possible in a realistic setting or whether, given the large number of free pinning sites available, a vortex would always re-pin before being able to move close to its neighbours, leading to a much more gradual, creep like, expulsion of vorticity.
In fact thermally activated creep is likely to play an important role in the dynamics of a superfluid neutron star \citep{Alpar84a, Link13}, and theories based on vortex creep have been successful in interpreting the post-glitch dynamics of many pulsars [see \citet{reviewHM} for a review of glitch relaxation models].
It is important to note that avalanches can also be driven by acoustic knock-on effects \citep{Lila12}. In this paper we do not consider this effect but only investigate avalanches due to proximity-knock on effects. 

We solve the equations of motion for a rigid vortex in a two-dimensional parabolic pinning potential, subject to mutual friction. We also consider the effect of bending and vortex tension and find that it alters the critical threshold for unpinning and sets the length-scale over which a vortex can remain rigid, with more rigid vortices being subject to weaker pinning forces \citep{Link09, Stefano}. Tension could thus play a role in re-pinning, as already noted by \citet{Link09}. Full three dimensional simulations are the natural next step to understand its impact in detail.
It is important to note that, however, the dynamics are insensitive to the exact value of the unpinning threshold. Rather, they are governed by how close the system is to the critical threshold, the strength of the mutual friction (parametrised by $\mathcal{R}$) and the initial position of the vortex in the pinning potential.

The main conclusion of this work is that there is a large section of parameter space where the mean free path of a vortex is larger than the typical vortex separation, and vortex-vortex interactions are possible, leading to proximity knock-on and avalanches. The main parameters that control the process are how close the system can come to the critical unpinning threshold, parametrised by $\Delta=(V_C-V^\n_y)/V_C$ and the strength of the mutual friction, parametrised by the dimensionless drag parameter $\mathcal{R}$. There is significant uncertainty regarding the exact value of $\mathcal{R}$ in the crust. If vortex motion past pinning centres can excite Kelvin waves the drag is likely to be quite strong, with values of $\mathcal{R}\approx 1$, while if vortex motion is slow the interaction is likely to proceed via interactions with sound waves in the lattice, leading to drag parameters as low as $\mathcal{R}\approx 10^{-10}$ \citep{PhononJones, Haskell12}. In general for weak drag vortices can always move far and give rise to avalanches. For stronger drag this depends critically on $\Delta$, i.e. on how close the system can get to the critical unpinning threshold. If thermal creep is highly efficient, as we could expect in younger, hotter stars, vortices randomly hopping to adjacent pinning sites may be able to spin the star down fast enough to maintain a large value of $\Delta$, and avalanches are not possible. However if creep is inefficient, which may be the case for colder stars, and if the superfluid drag is indeed weak ($\mathcal{R}\approx 10^{-10}$), the superfluid cannot spin-down fast enough and therefore must expel vorticity via discrete avalanche events. We find that even for relatively strong damping ($10^{-5}\lesssim \mathcal{R}\lesssim 1$) avalanches are possible once $\Delta\lesssim 0.05$ (i.e. once the lag is less than approximately $5\%$ of the critical lag), if the vortex is subject to the same drag inside and outside of the range of the pinning potential. Somewhat counter-intuitively the situation is worse if the drag is much weaker outside the pinning potential. In this case avalanches are only possible if the lag is less than approximately $0.1\%$ of the critical lag. 
Given that the parameters controlling repinning depend on temperature and density, it is possible that the same star can switch from being in a self organised critical state and expelling vorticity via avalanches, to a creep state in which voracity is expelled gradually. For example, as a star ages and cools, thermal creep becomes less efficient, decreasing $\Delta$ and increasing the probability of system spanning avalanches and large glitches. In fact we use micro physically motivated values for the drag parameter $\mathcal{R}$ and pinning force, obtained from \citet{Haskell12}, and show that avalanches are more likely in the higher density regions of the crust, which are the regions in which the pinning force is strongest. It is thus possible that physical differences between neutron stars, such as differences in mass and temperature, may explain the different glitching behaviours observed in the pulsar population.

The mean free path calculation in this paper assumes that $\Delta$ is determined exogenously and is uniform over the microscopic scales of interest. In reality, however, $\Delta$ is determined endogenously by the vortex motion itself. For example, if it is relatively high somewhere in the crust, vortices tend to pin preferentially near that location, increasing $\Delta$ locally in order to maintain $\mathcal{V}_y/\mathcal{A}$ near the unpinning threshold. In other words, the system continually self-adjusts to `erase' underlying `bumps' in the pinning landscape and preserve a state of marginal stability. Such SOC behaviour is analogous to a driven sandpile fluctuating about its critical slope. It is known to create long-range correlations in the system, which lead to power-law statistics for avalanche sizes and make system-spanning avalanches geometrically rather than exponentially rare \citep{Jensen98}. Long-range correlations in the variable $\Delta$ (or equivalently $\mathcal{V}_y/\mathcal{A}$) affect the repinning mean free path greatly. We do not include them in our calculation of $\lambda$ because we are not aware of an analytic method for doing so, short of re-running the Gross-Pitaevskii simulations in \citet{Lila11} in a much larger box. Hence our results for $\lambda$ do not apply reliably in the regime $\lambda \gg a$, as noticed previously (hence the cut-off at $\lambda=a$ that we have adopted). However the results are valid for $\lambda\lesssim a$, because the self-adjustment process is mediated by vortex motion and occurs on the inter-vortex scale $a$, not the inter-pinning-site scale. Hence our calculation of $\lambda$ is adequate for answering whether a vortex moves far enough before repinning to knock on its neighbour: the aim of this paper.

\section*{Acknowledgments}

BH acknowledges the support of the Australian Research Council (ARC) via a Discovery Early Career Researcher Award (DECRA) Fellowship. Am acknowledges the support of ARC Discovery Projects grant DP110103347.

\bibliographystyle{mn2e}
\bibliography{glitchrev}
\bsp

\appendix

\section{Curved vortices}
\label{app1}

\begin{figure}
\centerline{\includegraphics[scale=0.4]{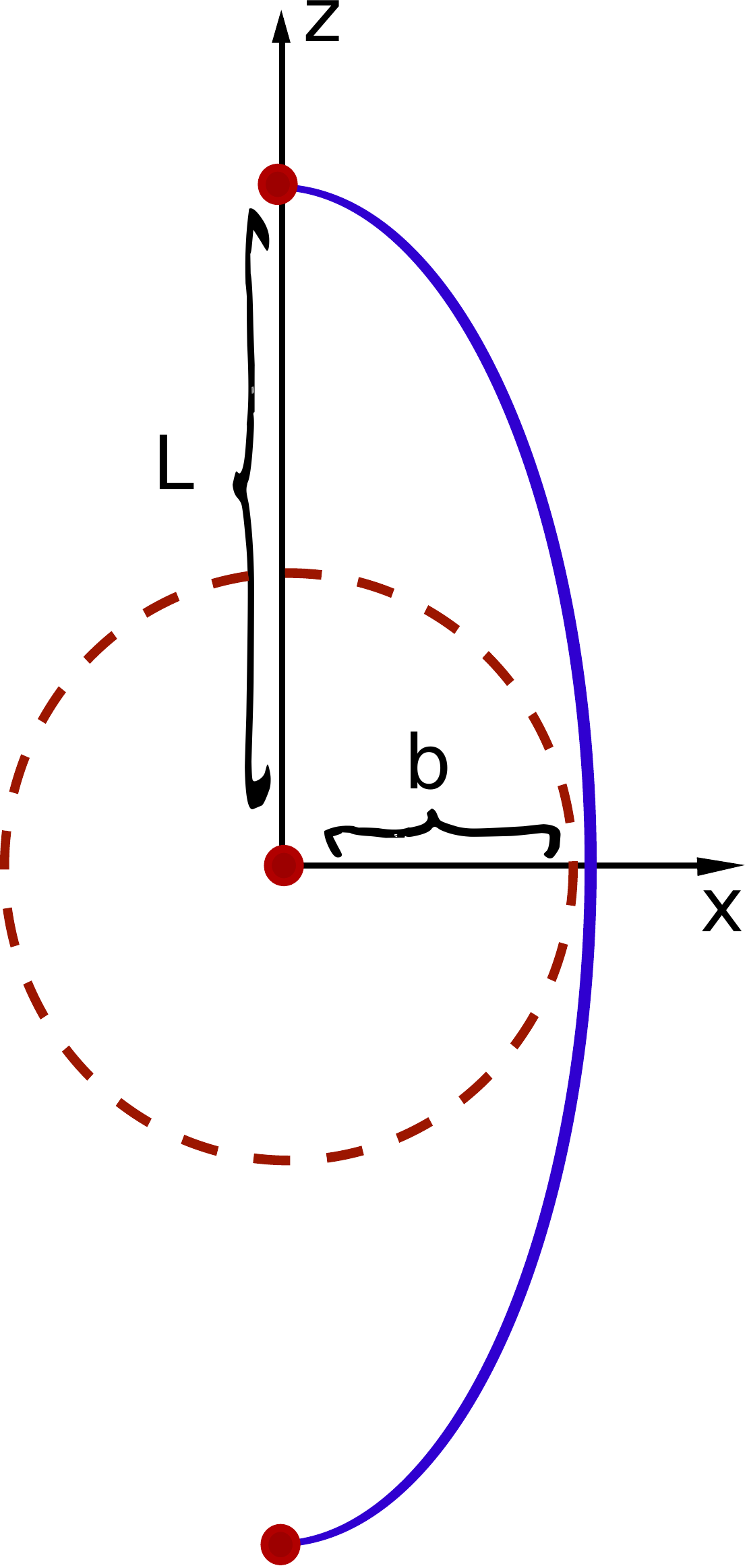}}
\caption{An out of scale representation of a curved vortex, pinned at its extrema, that has been displaced out of a pinning site at the origin along the $y$ axis by a distance $b$, over a length scale $L$. In a realistic neutron star one has $L\gg b$ and many more pinning sites, as discussed in the text. We approximate the configuration as an ellipse \citep{SoninShape}. Red dots along the $z$ axis represent (out of scale) pinning sites at the extrema of the vortex. The dashed circle represents the extent of the pinning potential. We assume that a vortex has just moved out of the potential, at a distance $b$ from the origin.}\label{ellipse}
\end{figure}

Consider a  pinned vortex that bends to free itself from one pinning bond. We take the same setup as previously, with the background neutron velocity in the $y$ direction. If the vortex moves out in the $x$ direction prior to unpinning, it will take approximately the shape of an ellipse in the $x-z$ plane. This configuration is shown in figure \ref{ellipse}, where $b$  is the semi-minor axis (and we assume that the vortex is at the edge of the range of the pinning potential), and $L$ is the semi-major axis, the length scale over which the vortex bends in the $z$ direction, which in this first example will be the distance to the next pinning site. For simplicity we neglect the effect of Mutual Friction in this section. The equations of motion for the vortex, from section \ref{teso}, take the form:
\be
v_i^L=V_i^\n+\epsilon_{ijk}\hat{k}^j\mathcal{F}^k-\hat{k}_i\hat{k}^j V_j^\n +\gamma_\n(\varepsilon)\nu\epsilon_{ijk}\hat{k}^j\hat{k}^p\nabla_p\hat{k}^k
\ee
where we have indicated as $V_i^\n$ the {\it background} superfluid neutron velocity (without the curvature induced contributions), and in the following we shall take $\gamma_\n=1$. Given that $\hat{k}_i$ has components only in the $x-z$ plane it is sufficient to consider the vortex line velocity in the x direction:
\be
v_y^L=V_y^\n+(\hat{k}^z\mathcal{F}^x-\hat{k}^x\mathcal{F}^z)+\nu[\hat{k}^z\hat{k}_p\nabla^p\hat{k}_x-\hat{k}^x\hat{k}_p\nabla^p\hat{k}_z]\label{generalfree1}
\ee
which, for $(x,z)=(b,0)$ (i.e. for a section of vortex that has just unpinned)  leads to:
\be
v_y^L=V_y^\n-\mathcal{A}b-\nu\frac{b}{L^2}\label{tensionMf1}
\ee
We can see from equation (\ref{tensionMf1}) that the tension acts in the same direction as the pinning force and tends to maintain the vortex straight. We can interpret this as a renormalisation of the strength of the pinning force, and define an effective pinning force:
\be
\tilde{\mathcal{F}}_i=-\tilde{\mathcal{A}}r_i
\ee
with
\be
\tilde{\mathcal{A}}=\mathcal{A}+\frac{\nu}{L^2}
\ee
The effective critical velocity $\bar{V}_{cT}$ for unpinning with tension, compared to the critical velocity $V_{cr}$ in the absence of tension, is thus
\be
\bar{V}_{cT}=V_{cr}+\frac{\nu R_{range}}{L^2}\label{newpin}
\ee
The first thing to note is that bending (and consequently unpinning) over the length scale of a single pinning site is prohibited by the tension, that in the case $L=R_{range}$ increases the effective pinning force $\tilde{\mathcal{F}}_i$ by several orders of magnitude, leading to critical velocities of the order of $\bar{V}_{cT}\approx 10^8$ cm/s, far greater than what is achievable in a neutron star. From equation (\ref{newpin}) we see that the critical velocity has a minimum in $L$, for a given pinning energy, i.e. for a fixed $\mathcal{A}$ or equivalently fixed $V_{cr}$. We can estimate that bending to unpin is possible (i.e. the tension will no longer tend to keep the vortex pinned and increase the critical unpinning velocity) if it occurs over length scales 
\be
L\gtrsim\sqrt{\frac{\nu R_{range}}{V_{cr}}}
\ee
For $R_{range}\approx R_\mathrm{WS}$, and if we approximate the critical velocity as $V_y^\n\approx E_p/(\rho_\n\kappa R_\mathrm{WS} L)$, this leads to $L\approx (T R_\mathrm{WS})/E_p\approx 10^3 R_\mathrm{WS}$ in the deep crust as estimated from energetics in (\ref{length}).

\section{Ordered lattice}
\label{app2}
Consider the case of an ordered square lattice, with pinning sites a distance $\xi R_{range}$ apart, where $\xi$ is a constant, and with the same pinning potential at every site. Given an initial entry angle $\theta$ of a vortex in a pinning site, we can calculate the exit angle of the vortex from \ref{repin} and then calculate the trajectory of the free vortex from (\ref{v1})-(\ref{v2}), to determine when it re-enters the range of at the pinning potential of an adjacent site, i.e. the updated angle $\theta$. 

\begin{figure}
\centerline{\includegraphics[scale=0.36]{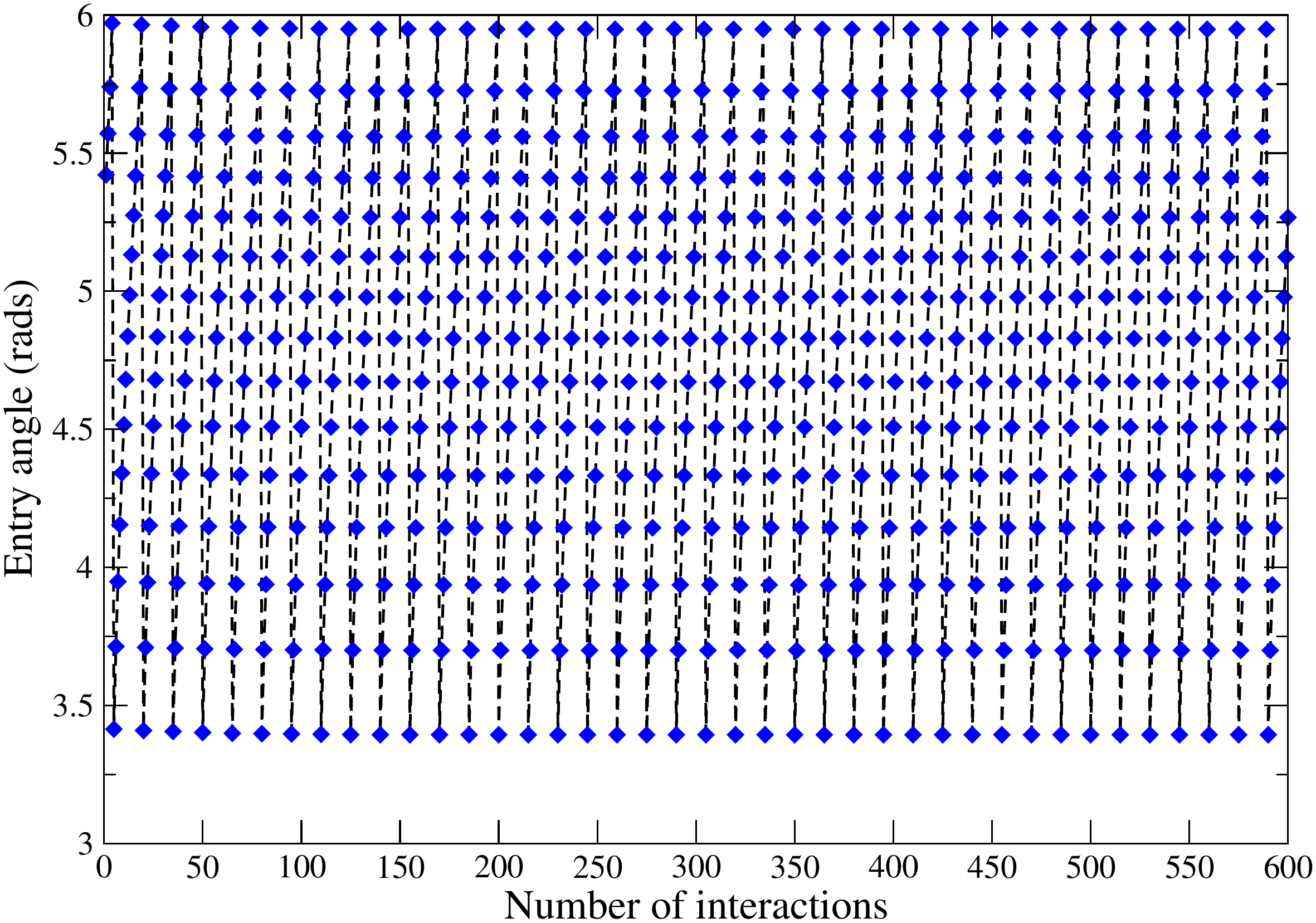}}
\caption{Re-entry angles $\theta$ of a vortex for repeated encounters with pinning sites in a square lattice, for $\mathcal{R}=0.5$, $\xi=2$ and $V_y/\mathcal{A}=0.99 R_{range}$. We see that all the possible range of  re-entry angles $\pi-\theta_d < \theta < 2\pi-\theta_d$  is scanned periodically and the vortex never repins. }\label{periodic}
\end{figure}

Successive  encounters are, naturally, correlated and the behaviour of the system is deterministic. For background flows far from the critical unpinning velocity (e.g. for a ratio $V_y/\mathcal{A}\lesssim R_{range}/2$) we find that in general the vortex repins after $\approx \mathcal{R}$ encounters. Closer to the critical unpinning velocity, however, we find several entry angles for which the vortex never repins, but rather can `hop' from one pinning site to the next, spanning the entire range of entry angles periodically. We show an example in figure \ref{periodic}, where we plot the re-entry angles of the vortex for each encounter for $\mathcal{R}=0.5$, $\xi=2$ and $V_y/\mathcal{A}=0.99 R_{range}$.  We see that the vortex will never repin as the allowed range of angles $\pi-\theta_d < \theta < 2\pi-\theta_d$ is scanned periodically. If the lattice is randomly spaced, it is arguable that encounters would no longer be correlated and the mean-free path approach described in this paper can be applied.

\label{lastpage}

\end{document}